\newcommand{\lsim}{\mathrel{\rlap{\lower4pt\hbox{\hskip1pt$\sim$}}
                   \raise1pt\hbox{$<$}}}
\newcommand{\gsim}{\mathrel{\rlap{\lower4pt\hbox{\hskip1pt$\sim$}}
                   \raise1pt\hbox{$>$}}}

\documentstyle[12pt]{article}
\begin{document}

\title{A General Supergravity Formalism for a
Naturally Flat Inflaton Potential
\footnote{Talk given at the Seventh Marcel Grossmann Meeting on
General Relativity, Stanford University, 24-29 July 1994, and based
on Ref.~1.} }
\author{Ewan D. Stewart
\thanks{e-mail address: eds@murasaki.scphys.kyoto-u.ac.jp} \\
Department of Physics \\ Kyoto University \\ Kyoto 606, Japan}
\maketitle
\begin{abstract}
A globally supersymmetric model of inflation will not work in a
generic supergravity theory because the higher order,
nonrenormalisable supergravity corrections destroy the flatness of
the inflaton's potential. In this talk I derive a form for the
K\"{a}hler potential which eliminates these corrections if
$ W = W_\varphi = \psi = 0 $ during inflation (where $W$ is the
superpotential, the inflaton $\in \varphi$, and $W_\psi \neq 0$).
I then point out that K\"{a}hler potentials of the required form
often occur in superstrings and that the target space duality
symmetries of superstrings often contain $R$-parities which would
make $ W = W_\varphi = 0 $ automatic for $ \psi = 0 $.
\end{abstract}
\vspace*{-87ex}
\hspace*{\fill}{\bf KUNS 1286}\hspace*{1.7em}\\
\hspace*{\fill}{hep-ph/9408302}
\thispagestyle{empty}
\setcounter{page}{0}
\newpage
\setcounter{page}{1}

\section{Introduction}

The approximate isotropy of the cosmic microwave background radiation
implies that the inflation that inflated the observable universe
beyond the Hubble radius must have occurred at an energy scale
$ V^{1/4} \leq 4 \times 10^{16}\,$GeV \cite{Liddle}.
Thus models of inflation should be constructed in the context of
supergravity \cite{susy}.

However, this immediately leads to a problem. The positive potential
energy $V>0$ required for inflation spontaneously breaks
supersymmetry.\footnote{After inflation $V$ disappears and so
supersymmetry is restored modulo whatever breaks supersymmetry in our
vacuum.} This would generally be expected to give an effective mass
squared $ m_{\rm soft}^2 \sim 8\pi V / m_{\rm Pl}^2 \sim H^2 $
to any would-be inflaton. But the effective mass of the inflaton must
be much less than the inflationary Hubble parameter $H$.

In this talk I will present a general formalism that solves this
problem.

\subsection{Notation and basic formulae}

I set $m_{\rm Pl}/\sqrt{8\pi} = 1$ throughout.
$\phi$ will represent a vector whose components $\phi^\alpha$ are
complex scalar fields, and subscript $\phi$ will denote the derivative
with respect to $\phi$, so for example $W_\phi$ represents the vector
with components $ \partial W / \partial \phi^\alpha $.

The scalar potential in a globally supersymmetric theory is
\begin{equation}
V = \left| W_\phi \right|^2 + D{\rm -term} \,,
\end{equation}
where the superpotential $W(\phi)$ is an analytic function of $\phi$.
The first term is called the $F$-term. For simplicity I will ignore
the $D$-term in this talk.

The $F$-term part of the scalar potential in a supergravity theory is
\begin{equation}
\label{V}
V = e^K \left[ \left( W_\phi + W K_\phi \right)
        K^{-1}_{\bar{\phi} \phi}
        \left( \bar{W}_{\bar{\phi}} + \bar{W} K_{\bar{\phi}} \right)
        - 3 |W|^2 \right] \,,
\end{equation}
where the K\"{a}hler potential $K(\phi,\bar{\phi})$ is a real function
of $\phi$ and its hermitian conjugate $\bar{\phi}$.

\section{The Problem}
\label{prob}

At any point in the space of scalar fields $\phi$ we can make a
combination of a K\"{a}hler transformation and a holomorphic field
redefinition such that $\phi=0$ at that point and, in the
neighbourhood of that point, the K\"{a}hler potential takes the form
\begin{equation}
K = \left| \phi \right|^2 + \ldots \,,
\end{equation}
where \ldots\ stand for higher order terms.
Then the scalar kinetic terms will be canonical at $\phi=0$ and,
from Eq.~(\ref{V}), the scalar potential will have the form
\begin{eqnarray}
V & = & e^{ \left| \phi \right|^2 + \ldots }
	\left\{ \left[ W_\phi + W \left( \bar{\phi} + \ldots \right)
	\right] \left( 1 + \ldots \right) \left[ \bar{W}_{\bar{\phi}}
	+ \bar{W} \left( \phi + \ldots \right) \right] - 3|W|^2 \right\}
	\,, \\
 & = & V|_{\phi=0} + V|_{\phi=0} \left| \phi \right|^2
	+ {\rm other\ terms} \,.
\end{eqnarray}
Thus at $\phi=0$ the exponential term gives a contribution $V$ to the
effective mass squared of {\em all\/} scalar fields. Therefore,
\begin{equation}
\frac{ V''}{V} = 1 + {\rm other\ terms} \,,
\end{equation}
where the prime denotes the derivative with respect to the canonically
normalised inflaton field. But $|V''/V| \ll 1$ is necessary for
inflation to work.
So a sucessful model of inflation must arrange for a cancellation
between the exponential term and the terms inside the curly brackets.
This will require fine tuning unless a symmetry is used to enforce it.

\section{A Solution}

Divide the vector of scalar fields $\phi$ into two separate vectors,
$\varphi$ and $\psi$, with the inflaton contained in $\varphi$:
\begin{equation}
\phi = (\varphi,\psi) \,,\;\;\;\; {\rm inflaton} \in \varphi \,.
\end{equation}
Assume the $R$-parity
\begin{equation}
\label{R}
\psi \rightarrow -\psi \,,\;\;\;\; \varphi \rightarrow \varphi
\,,\;\;\;\;W \rightarrow -W \,,\;\;\;\; K \rightarrow K \,,
\end{equation}
and that during inflation
\begin{equation}
\psi = 0
\end{equation}
(a natural value since the necessary condition $V_\psi = 0$ is then
guaranteed by the $R$-parity). Then the $R$-parity ensures that during
inflation
\begin{equation}
W = W_\varphi = 0 \,.
\end{equation}
Thus the scalar potential Eq.~(\ref{V}) simplifies to
\begin{equation}
V = e^K W_\psi K^{-1}_{\bar{\psi} \psi} \bar{W}_{\bar{\psi}} \,.
\end{equation}
Now it becomes possible to choose a form for the K\"{a}hler
potential that cancels the inflaton dependent corrections to the
global supersymmetric potential in a natural way.

Expanding the $R$-parity invariant K\"{a}hler potential about
$\psi = 0$ gives
\begin{equation}
K = A(\varphi,\bar{\varphi}) + \bar{\psi}\, B(\varphi,\bar{\varphi})
	\,\psi + {\cal O} \left( \psi^2 , \bar{\psi}^2 \right) \,,
\end{equation}
where $A$ is real and $B$ is hermitian. Therefore
\begin{equation}
V = e^A W_\psi B^{-1} \bar{W}_{\bar{\psi}} \,,
\end{equation}
and so to eliminate the inflaton dependent corrections to the global
supersymmetric potential we require
\begin{equation}
B^{-1} = f(\varphi,\bar{\varphi}) \, C^{-1}(\chi,\bar{\chi}) \,,
\end{equation}
and
\begin{equation}
A = - \ln f(\varphi,\bar{\varphi}) + g(\chi,\bar{\chi}) \,,
\end{equation}
where $f$ and $g$ are real functions, $C$ is a hermitian matrix, and
$\chi$ are non-inflaton $\varphi$ fields. This gives the inflationary
potential
\begin{equation}
V = e^{ g(\chi,\bar{\chi}) }
	W_\psi\, C^{-1}(\chi,\bar{\chi}) \,\bar{W}_{\bar{\psi}} \,,
\end{equation}
and the K\"{a}hler potential is required to have the general form
\begin{eqnarray}
K & = & - \ln f(\varphi,\bar{\varphi})
	+ \frac{ \bar{\psi}\, C(\chi,\bar{\chi}) \,\psi }
	{ f(\varphi,\bar{\varphi}) } + g(\chi,\bar{\chi})
	+ {\cal O} \left( \psi^2 , \bar{\psi}^2 \right) \,, \\
\label{K}
 & = & - \ln \left[ f(\varphi,\bar{\varphi})
	- \bar{\psi}\, C(\chi,\bar{\chi}) \,\psi \right]
+ g(\chi,\bar{\chi}) + {\cal O} \left( \psi^2 , \bar{\psi}^2 \right)
	\,.
\end{eqnarray}

\section{Superstring Examples}

\subsection{Orbifold compactifications}

The K\"{a}hler potential of the untwisted sector of the low-energy
effective supergravity theory derived from orbifold compactification
of superstrings always contains \cite{Ferrara}
\begin{equation}
K = - \ln \left( S + \bar{S} \right)
- \sum_{i=1}^{3} \ln \left( T_i + \bar{T}_i
- \left| \phi_i \right|^2 \right) \,,
\end{equation}
where $S$ is the dilaton, $T_i$ are the untwisted moduli associated
with the radii of compactification, and $\phi_i$ are the untwisted
matter fields associated with $T_i$.
Now if we divide the scalar fields into $\varphi$, $\psi$ and $\chi$
fields as follows
\begin{eqnarray}
T_1 & \in & \varphi \,, \\
\phi_1 & \in & \psi \,, \\
S \,,\, T_2 \,,\, T_3 \,,\, \phi_2 \;{\rm and}\; \phi_3 & \in &
\chi \subset \varphi \,,
\end{eqnarray}
then we get a K\"{a}hler potential of the required form
[Eq.~(\ref{K})]
\begin{equation}
K = - \ln \left( \varphi + \bar{\varphi} - \left| \psi \right|^2
	\right) + g(\chi,\bar{\chi}) \,,
\end{equation}
and the target space duality symmetries,
\begin{equation}
T_i \rightarrow \frac{a_i T_i - i b_i}{i c_i T_i + d_i} \,,\;\;\;\;
\phi_i \rightarrow \frac{\phi_i}{i c_i T_i + d_i} \,,\;\;\;\;
a_i d_i - b_i c_i = 1 \,,
\end{equation}
contain the desired $R$-parity [Eq.~(\ref{R})] on setting
$ b_i = c_i = 0 $, $ a_1 = d_1 = -1 $ and
$ a_2 = a_3 = d_2 = d_3 = 1 $.

\subsection{Fermionic four-dimensional string models}

The K\"{a}hler potential of the untwisted sector of the revamped
flipped SU(5) model \cite{flipped} is \cite{Lopez}
\begin{eqnarray}
\lefteqn{ K = -\ln \left( 1 - \left| \Phi_1 \right|^2
- \left| \Phi_{23} \right|^2 - \left| \Phi_{\overline{23}} \right|^2
- \left| h_1 \right|^2 - \left| h_{\overline{1}} \right|^2
+ \frac{1}{4} \left| \Phi_{1}^{2} + 2 \Phi_{23} \Phi_{\overline{23}}
	+ 2 h_1 h_{\overline{1}} \right|^2 \right) } \nonumber \\
 & & \mbox{} - \ln \left( 1 - \left| \Phi_2 \right|^2
- \left| \Phi_{31} \right|^2 - \left| \Phi_{\overline{31}} \right|^2
- \left| h_2 \right|^2 - \left| h_{\overline{2}} \right|^2
+ \frac{1}{4} \left| \Phi_{2}^{2} + 2 \Phi_{31} \Phi_{\overline{31}}
	+ 2 h_2 h_{\overline{2}} \right|^2 \right) \nonumber \\
 & & \mbox{} - \ln \left( 1 - \left| \Phi_4 \right|^2
- \left| \Phi_5 \right|^2 - \left| \Phi_3 \right|^2
- \left| \Phi_{12} \right|^2 - \left| \Phi_{\overline{12}} \right|^2
- \left| h_3 \right|^2 - \left| h_{\overline{3}} \right|^2
\right. \nonumber \\
 & & \left. \hspace{3em} \mbox{} + \frac{1}{4} \left| \Phi_{4}^{2}
	+ \Phi_{5}^{2} + \Phi_{3}^{2} + 2 \Phi_{12} \Phi_{\overline{12}}
	+ 2 h_3 h_{\overline{3}} \right|^2 \right) \,.
\end{eqnarray}
Now if we divide the fields as follows
\begin{eqnarray}
\Phi_4 \;{\rm and}\; \Phi_5 & \in & \varphi \,, \\
\Phi_3 \,,\, \Phi_{12} \,,\, \Phi_{\overline{12}} \,,\, h_3
	\;{\rm and}\; h_{\overline{3}} & \in & \psi \,, \\
\Phi_1 \,,\, \Phi_2 \,,\, \Phi_{23} \,,\, \Phi_{\overline{23}} \,,\,
	\Phi_{31} \,,\, \Phi_{\overline{31}} \,,\, h_1 \,,\,
	h_{\overline{1}} \,,\, h_2 \;{\rm and}\; h_{\overline{2}} & \in &
	\chi \subset \varphi \,,
\end{eqnarray}
then we get a K\"{a}hler potential of the required form
[Eq.~(\ref{K})]
\begin{equation}
K = - \ln \left( 1 - \left| \varphi \right|^2
	+ \frac{1}{4} \left| \varphi^{\rm T} \varphi \right|^2
	- \left| \psi \right|^2 \right) + g(\chi,\bar{\chi})
+ {\cal O} \left( \psi^2 , \bar{\psi}^2 \right) \,,
\end{equation}
and the target space duality symmetries \cite{Lopez} contain the
desired $R$-parity [Eq.~(\ref{R})].

\subsection{More orbifold compactifications}

A K\"{a}hler potential of the form
\begin{equation}
K = - \ln \left[
	\left( A_1 + \bar{A}_1 \right) \left( A_2 + \bar{A}_2 \right)
	- \left( A_3 + \bar{A}_4 \right) \left( A_4 + \bar{A}_3 \right)
	\right]
\end{equation}
often occurs in orbifold compactifications
\cite{Ferrara,Cvetic,Cardoso} and is of the required form.

\subsection{Calabi-Yau compactifications}

K\"{a}hler potentials of the form
\begin{equation}
K = - \ln \left( 1 - |N|^2 - |C|^2 \right)
	+ {\cal O} \left( C^2 , \bar{C}^2 \right)
\end{equation}
occur for subspaces of enhanced symmetry of the moduli space of a
simple Calabi-Yau manifold \cite{Dixon}. They are of the required
form [Eq.~(\ref{K})].

\section{Summary}

A globally supersymmetric model of inflation will not work in a
generic supergravity theory because the higher order,
nonrenormalisable supergravity corrections destroy the flatness of
the inflaton's potential. In this talk I have derived a form for the
K\"{a}hler potential which eliminates these corrections if
$ W = W_\varphi = \psi = 0 $ during inflation (where $W$ is the
superpotential, the inflaton $\in \varphi$, and $W_\psi \neq 0$). It
is encouraging that K\"{a}hler potentials of the required form often
occur in superstrings and that the target space duality symmetries of
superstrings often contain $R$-parities which would make
$ W = W_\varphi = 0 $ automatic for $ \psi = 0 $.

\subsection*{Acknowledgements}
I thank D. H. Lyth for his help in simplifying the presentation of
this work.
I am supported by a JSPS Postdoctoral Fellowship and this work was
supported by Monbusho Grant-in-Aid for Encouragement of Young
Scientists No.\ 92062.

\end{document}